# Disclosing antiferromagnetism in tetragonal $Cr_2O_3$ by electrical measurements

M. Asa[1], C. Autieri[2], C. Barone[3], C. Mauro[3], S. Picozzi[4], S. Pagano[3] and M. Cantoni[1,a]

[1] *Dipartimento di Fisica, Politecnico di Milano, Via G. Colombo 81, 20133 Milano, Italy*

[2] *International Research Centre MagTop, Institute of Physics, Polish Academy of Sciences, Aleja Lotników 32/46, PL-02668 Warsaw, Poland*

[3] *Dipartimento di Fisica "E.R. Caianiello" and CNR-SPIN Salerno, Università di Salerno, 84084 Fisciano, Salerno, Italy*

[4] *Consiglio Nazionale delle Ricerche, Istituto Superconduttori, Materiali Innovativi e Dispositivi (CNR-SPIN), c/o Università G. D'Annunzio, I-66100 Chieti, Italy*

## ABSTRACT

The tetragonal phase of chromium (III) oxide, although unstable in the bulk, can be synthesized in epitaxial heterostructures. Theoretical investigation by density functional theory predicts an antiferromagnetic ground state for this compound. We demonstrate experimentally antiferromagnetism up to 40 K in ultrathin films of t-$Cr_2O_3$ by electrical measurements exploiting interface effect within a neighboring ultrathin Pt layer. We show that magnetotransport in Pt is affected by both spin-Hall magnetoresistance and magnetic proximity effect while we exclude any role of magnetism for the low-temperature resistance anomaly observed in Pt.

---

[a] Author to whom correspondence should be addressed: matteo.cantoni@polimi.it

**PACS**:

# I. INTRODUCTION

Although antiferromagnets (AFMs) were thought from their discoverer himself "not to have any practical application" in 1970 [1], they currently represent a fundamental building block of magnetic memories and sensors, based on the spin-valve structure exploiting magnetoresistive effects (GMR, TMR). [2] Beside this consolidate technological application, which still relies on the presence ferromagnetic materials, the concept of purely antiferromagnetic devices is gaining more and more interest in the last years within the emerging field of antiferromagnet spintronics [3,4]. In fact, a magnetic storage device without ferromagnetic elements, replaced instead by antiferromagnetic ones, would allow higher scalability, robustness, and writing speed.

A key task is to achieve in AFM-based devices a comparable level of control as in ferromagnetic ones. To this scope, electrical methods to manipulate and/or readout the state of AFM materials were developed and demonstrated in AFM compounds like FeRh, [5] CuMnAs, [6] and Mn2Au. [7] In these systems the spin configuration of the antiferromagnet can be detected thanks to the mechanism of Anisotropic MagnetoResistance (AMR) which, ubiquitous in ferromagnetic metals, can also be observed in some metallic antiferromagnets. For insulating antiferromagnets, instead, different approaches are needed as no current can flow in them. Two possible solutions have been proposed in this regard, exploiting interfaces of the AFM with suitable metals: i) to measure the spin-Hall magnetoresistance (SMR) in heavy metals interfacing AFMs, [8–16] and ii) detect the proximity induced magnetization in a neighboring layer that would be normally non-magnetic (magnetic proximity effect, MPE). [17,18] While the former effect has been shown to cause minor changes of longitudinal resistivity, Hall measurements in AFMs interfaced with Pt have been reported to display a sizable signal that has been attributed to MPE. [17,18]

Interface effects can be exploited as a tool to investigate the magnetic properties in new or unknown materials. In this regard, we consider in this paper the case of tetragonal chromium (III) oxide, t-$Cr_2O_3$, an unstable phase of chromium oxide which can be stabilized exploiting epitaxial growth on suitable substrates [19,20].

We first report the results from first-principles density functional calculations indicating an antiferromagnetic character for t-$Cr_2O_3$. Then, by measuring the transport properties in an ultrathin Pt film deposited on t-$Cr_2O_3$ as a function of temperature and external field, we reveal the presence of interfacial effects. Electrical measurements allow to demonstrate unambiguously the antiferromagnetic ground state of this oxide compound, as predicted also by *ab-initio* calculations. Monitoring the thermal behavior of transport in Pt both in longitudinal and transverse configurations, exploiting a general approach that can be used in principle with any antiferromagnetic compound, we establish the Neèl temperature of ultrathin t-$Cr_2O_3$ films to be close to 40 K. Finally, we provide some evidences indicating that both SMR and MPE coexist at this interface and both produce measurable effects on the electronic transport.

## II. COMPUTATIONAL DETAILS AND EXPERIMENTAL SETUP

First-principles density functional theory calculations were performed within the generalized gradient approximation (GGA) by using the plane-wave VASP package [21]. Additional information about the computational framework can be found in reference [20]. The Hubbard U effects on the Cr sites were included within the GGA + U approach [22], using the rotational invariant scheme [23]. Using a value of the Hund coupling constant $J_H = 0.15U$, we performed the calculations for U = 6 eV since it is a value generally used for correlated compounds presenting a reduced electron bandwidth. To prove that, we compare the $t_{2g}$ bandwidth of the t-$Cr_2O_3$ with other Cr-compounds with octahedral crystal field. Indeed, the t-$Cr_2O_3$ shows a $t_{2g}$ bandwidth of 2.5 eV that is small compared with the bandwidth of 3.5 eV for the CrAs [24] or the bandwidth of 5.0 eV for the elemental Cr [25]. A $12 \times 4 \times 6$ k-point Monkhorst-Pack grid was used for the Brillouin zone sampling for the 10 atoms supercell [26], while a $6 \times 4 \times 6$ k-point grid was used when we doubled the cell along the a-axis in order to determine the magnetic exchanges.

Pt/t-$Cr_2O_3$/$BaTiO_3$ thin films were grown on commercial SrTiO3 substrates by combining pulsed laser deposition (PLD), molecular beam epitaxy (MBE), and thermal treatments. As described in detail in [20], single-crystalline $BaTiO_3$ films are first grown by PLD. Then, ultrathin metallic Cr (thickness = 2 nm) is deposited by MBE in ultrahigh vacuum regime (pressure<$10^{-9}$ mbar) on top of it and then annealed at 800 K for 30 minutes. The temperature promotes diffusion of oxygen from $BaTiO_3$ towards the overlayer [27], forming an epitaxial film of chromium (III) oxide about 3 nm-thick (the layer grows because of the oxygen inclusion) which is stable in the defective rocksalt configuration (t-$Cr_2O_3$). With this method, a single t-$Cr_2O_3$ phase in the film has been demonstrated by electron and photoelectron diffraction. In the same way, a uniform Cr2+ oxidation state has been proven within the t-$Cr_2O_3$ layer by scanning transmission electron microscopy and electron-energy loss spectroscopy (STEM-EELS) with atomic layer resolution. [20] A 2 nm thick Pt layer is finally deposited by MBE with a deposition rate of 0.7 nm/min. The whole process is carried without breaking the vacuum in a cluster tool allowing for both deposition and sample preparation steps [28]. Subsequently, devices for electrical characterization were defined by a two-step process of optical lithography and ion milling in the form of as Hall crosses with a lateral dimension of 50 μm.

Longitudinal and transverse resistance measurements were performed as a function of temperature and magnetic field using a current-source/nanovoltmeter pair Keithley 6221/2182A operating in Delta Mode. A relay switching matrix was used to acquire both transverse and longitudinal measurements at the same time. The measurement setup was connected to a cryogenic system with superconducting coils able to provide a magnetic field up to 7.5 T. For measurements at remanence, the hysteresis of the superconducting coils has been calibrated and compensated within an error of ±1 Oe.

Pristine t-$Cr_2O_3$ ultrathin films on $BaTiO_3$ (i.e. without the top Pt layer) were electrically characterized displaying a negligible conductivity even with macroscopic contact pads. Therefore, we exclude any contribution on the electronic transport form the oxygen-deficient layer within $BaTiO_3$.

Additional characterization of the transport properties was made by electric noise spectroscopy measurements. A dedicated measurement setup, consisting in a 8-300 K closed-cycle refrigerator (Janis Research), a low-noise dc current source (Keithley 220), a low-noise preamplifier (Signal Recovery 5113), and a dynamic signal analyzer (HP35670A), has been employed [29]. Unwanted contact noise contributions were removed by resorting to a specific procedure, based on a sequence of two- and four-probe measurements [30]. The instrumental background noise level is $1.4 \times 10^{-17}$ $V^2$/Hz.

## III. RESULTS

### A. Ab-initio calculations of magnetic properties in t-$Cr_2O_3$

As tetragonal Chromium (III) oxide has been synthesized only recently in epitaxial heterostructures, [19,20] its magnetic properties are mostly unknown. To unveil the magnetic character of this compound, density functional theory calculations were performed on the structure determined in ref. [20] for t-$Cr_2O_3$ grown on $BaTiO_3$ and reported in Figure 1(a).

According to our calculations, the magnetic moment of the Cr atoms in t-$Cr_2O_3$ is 2.9 $\mu_B$, very close to 3 $\mu_B$ that is the atomic limit of the spin moment of $Cr^{3+}$. We examined all the possible collinear spin configurations. The magnetic ground state is represented in Figure 1(a)

We estimated the magnetic exchanges for the three inequivalent first-neighbors of the Cr sublattice. We described the Cr-Cr exchange interactions in terms of the classical Heisenberg Hamiltonian with spin S=1

$$H = \sum_{<i,j>} J_{ij}\vec{S_i} \cdot \vec{S_j}$$

where i and j run over the Cr-site. The symbol $<i,j>$ specifies that we take in account just the first-neighbors. In our convention both $J_{ij}$ and $J_{ji}$ must be considered.

The system presents magnetic exchange along the a-axis ($J_a$), along the b-axis ($J_b$), and atoms with different z coordinate ($J_c$), as shown in Figure 1(b). We calculated the total energy of the three possible antiferromagnetic collinear phases with zero net magnetic moment since these magnetic phases are the closest to the ground state. From these energies we estimated the magnetic exchanges constants $J_b$ and $J_c$. We doubled the cell along the a-axis and we repeat the procedure to obtain $J_a$.

We obtained a ferromagnetic exchange along the a-axis, whereas the magnetic exchanges along the $J_b$ and $J_c$ are antiferromagnetic. From a quantitative point of view, $J_b$=26.1 meV is the dominant magnetic exchange, while the other contributions result $J_a$=-3.0 meV and $J_c$=1.6 meV.

Starting from the previous Heisenberg Hamiltonian, it is possible to derive the critical temperature in the mean field approximation as $T_C^{MFA} = \frac{2}{3k_B}\sum_{i=1}^{n} J_{0i}$ [31], where n is the number of first-neighbors and $J_{0i}$ the magnetic exchange between the Cr atom and the i-th neighbor. Considering the numerical values of the magnetic exchange constants and the coordination number, the critical temperature in mean field approximation results 205 K.

### B. Electrical properties of Pt on t-Cr$_2$O$_3$

#### *1. Temperature dependence of resistivity*

The longitudinal resistivity $\rho_{xx}$ of Pt grown on t-Cr$_2$O$_3$ from 6 to 300 K, obtained in Van der Pauw geometry, [32] is shown in Figure 2(a). As typically noticed in ultrathin Pt films, the resistivity is sensibly larger at room temperature than the bulk value $\rho$ = 11.1 μΩ cm. The ordinary temperature dependence for good metals is observed with a positive coefficient of resistance d$\rho$/dT over most of the range inspected. Nevertheless, at low temperature an anomaly is found (see inset in Figure 2a) and the resistivity grows back again with a logarithmic trend below 20 K. As noted in similar systems [33], two possible mechanisms can account for this behavior: weak localization [34] and scattering with magnetic impurities (Kondo effect). [35]

Since Kondo effect has been recently observed in Pt made ferromagnetic by the application of strong electric fields [36], its presence in our Pt/t-Cr$_2$O$_3$ would indicate proximity induced magnetization in Pt made ferromagnetic by the underlying t-Cr$_2$O$_3$ layer (magnetic proximity effect).Therefore, a systematic study has been carried in this system in order to verify the physical origin of the low-temperature anomaly, exploiting the spectroscopy of charge carriers fluctuations. This analysis is essentially based on the measurement of the voltage-spectral density $S_V$ generated by the device under test when biased by a constant DC current. The frequency dependence of $S_V$ is shown in Figures 2(b) and 2(c) for different bias current values and at temperatures of 10 and 50 K, respectively. Two distinct noise components can be identified. The first is the 1/f noise, characteristic of the low-frequency region and associated to electrical conductivity fluctuations [37]. The second one is a frequency-independent "white noise" component, characteristic of the high frequency region of the spectrum and, usually, given by the sample thermal noise added to the instrumental background contribution.

Useful information on the transport mechanisms in action can be extracted by studying the 1/f noise amplitude and its dependence on external parameters, such as temperature, bias current, magnetic field, etc. In particular, it is well-known that standard resistance fluctuations, usually dominant in metallic systems, are characterized by a quadratic current dependence of the 1/f noise for all temperatures [37,38]. In systems exhibiting nonequilibrium universal conductance fluctuations, associated to weak-localization effects, a linear bias dependence of the 1/f noise is observed at low temperatures [39,40]. More recently, the evidence of Kondo effect due to magnetic impurities in granular aluminum has been connected to magnetic-dependent resistance fluctuations [41].

By analyzing the measured noise spectra, it is possible to derive the dependence of the 1/f amplitude on the bias current. In Figures 2(d) and 2(e), it is clearly shown that at low temperatures (10 K) such dependence is almost linear, while at higher temperatures (50 K) it is substantially quadratic. A quantitative information can be obtained by fitting the experimental data with a second-order polynomial

$$S_V = a_2 I^2 + a_1 I + a_0,$$

where $a_2$, $a_1$, and $a_0$ are the quadratic, linear, and constant noise parameters. The temperature dependence of $a_2$ and $a_1$, obtained from best fitting procedure, is shown in Figure 2f. A crossover between a high-temperature quadratic behavior and a low-temperature linear one is evident at T > 30 K. This finding is a clear evidence of the occurrence of a weak-localization transport mechanism at low temperatures, as already observed in a variety of materials [39,40].

As noted before, the DC resistivity shows a deviation from an almost linear behavior at T < 40 K followed by an upturn below 20 K, well fitted by a logarithmic temperature dependence, that falls in the weak localization region. Since the DC measurements alone cannot allow to discriminate between weak localization and Kondo effect as the cause of the resistivity anomaly, the noise analysis gives a clear indication in favor of the former. Additionally, no magnetic field dependence of the noise amplitude, which was found to be a signature of Kondo effect in granular aluminum oxide systems [41,42], has been observed for fields up to 1000 Gauss.

## *2. Magnetoresistance and Hall effect*

The magnetic properties of t-$Cr_2O_3$ have been further investigated by studying magnetoresistance and Hall transport in the Pt layer as a function of temperature. The results are reported in Figure 3.

Low-field longitudinal magnetoresistance (MR) calculated as $\frac{R_{xx}(B)-R_{xx}(0)}{R_{xx}(0)}$ is shown in Figure 3(a-c) for different relative orientations of the magnetic field **B** and the current **j**. As a first general observation for all the three cases, we note that the MR monotonously decreases in absolute value when lowering the sample temperature, eventually displaying an almost flat response for T ≤ 40 K. This is a first signature of a magnetic phase transition from the paramagnetic to the antiferromagnetic state in t-$Cr_2O_3$, and a similar behavior can also be observed at the phase transition of $Cr_2O_3$ in the corundum structure [11].

In fact, the MR and its thermal behavior above 40 K can be largely explained in the context of spin-Hall magnetoresistance for a paramagnetic material. We notice that a positive MR is observed for **B**∥**j** (Figure 3(a)) and **B**⊥**j** with **B** normal to the sample surface (Figure 3(c)), while negative MR is found when **B**⊥**j** with **B** in plane (Figure 3(b)). This is consistent with the SMR expected on a paramagnet where the spins are aligned with the applied field [43].

We also observe that the MR ratio rapidly increases with temperature just above the transition and then flattens out at higher temperature. For instance, the curves for T = 150 K and T = 200 K are very close in graphs 3(a-c) indicating that the temperature trend has almost saturated. As a matter of fact, in this range we are already deep in the paramagnetic phase of the material and we do not expect anymore a substantial dependence of SMR on temperature.

In the antiferromagnetic phase (T ≤ 40 K), instead, low magnetic fields cannot perturb the spin configuration because of the negligible magnetic susceptibility of the material, suppressing any magnetoresistive effect below the resolution limit.

Hall measurements at several temperatures are reported in Figure 3(d). Geometrical offsets were compensated by averaging for each point two separate measurements with current and voltage probes exchanged [44].

At low temperatures (T ≤ 40 K), only the linear component of the transverse resistance is observed. This can be unambiguously ascribed to the ordinary Hall effect in Pt. A positive sign of the Hall coefficient has already been reported for Pt ultrathin films thinner than 3 nm, as in our case. [36,45].

Above 40 K, instead, a non-linear anomalous component arises as a result of the out of plane component of magnetization of the underlying t-$Cr_2O_3$ in the paramagnetic state.

These results are consistent with Hall measurements across the antiferromagnetic phase transition at 311 K of corundum α-$Cr_2O_3$ [17] and, together with magnetoresistance measurements, corroborate the magnetic phase transition in ultrathin t-$Cr_2O_3$ around 40 K.

We note that, differently from the case of MR, the AHE signal keeps increasing with temperature also between T = 200 K and T = 300 K. In general, AHE evolves with temperature independently of magnetization (that we expect to be constant under the same applied field in this temperature range) and we ascribe this behavior, instead, to the temperature dependence of transport coefficients of the Pt layer.

A final indication of antiferromagnetic ordering in t-Cr$_2$O$_3$ below 40 K is given by the measurement of anomalous Hall effect at remanence following field cooling process. In this experiment Pt/t-Cr$_2$O$_3$ is cooled through the phase transition with a magnetic field of ±7.5 T applied out-of-plane (field cooling). When the magnetic field is removed, a non-zero signal in transverse resistance in observed. Its value depends only on the sign of the magnetic field applied during the field cooling. Figure 4 reports the difference in transverse resistance R$_{xy}$ between states written with +7.5 T and -7.5 T field cooling, whereas the sample is warmed up in zero external field. The difference between the states, indication of a memory effect frozen in the antiferromagnet spin structure, is apparent only up to 37±5 K. We note that field cooling at 7.5 T already guarantees the maximum imprinted magnetic state and identical signals where collected after field cooling at 3 T. This non-zero signal in the AFM phase could be surprising coming from a G-type antiferromagnet where the moments are compensated within each layer. A possible origin could be the intrinsic small magnetic moment of antiferromagnetic domain walls [46], which would present a preferential orientation as a consequence of the field cooling process, frozen in the antiferromagnetic lattice at first and then destroyed (the signal becomes zero, within the error bar) when the system becomes paramagnetic. As some of the authors have already shown in the case of Pt/Cr [47], the point at which the remnant AHE goes to zero corresponds to the Neèl temperature of the antiferromagnet. Considering the experimental error bar, this measurement confirms the previous indications of antiferromagnetic character of t-Cr$_2$O$_3$ with a transition at about 40 K. We note that this value is significantly smaller than that of bulk α-Cr$_2$O$_3$ (311 K). The transition temperature predicted by ab-initio calculations for the tetragonal phase (205 K) is already smaller than the one of the corundum phase just because the different structure determines a lower magnetic coupling between Cr atoms of the two structures. Moreover, this theoretical critical temperature calculated in mean field approximation is generally overestimated. As a matter of fact, this method works quite well in the case of high-dimensions and large connectivity. When the magnetic sublattice presents low connectivity as the Cr sublattice in t-$Cr_2O_3$, instead, the experimental critical temperature can be lower, as already demonstrated in other low-connectivity oxides [48,49]. Additionally, defects and inhomogeneity, which are related to weak localization experimentally observed through noise spectroscopy, can lower the critical transition temperature from the ideal case considered in theoretical calculations.

Finally, a reduction of the transition temperature in thin films has been observed for α-$Cr_2O_3$, films when thickness is comparable with the spin correlation length. [50] A similar finite-size scaling effect could also be present in our ultrathin films of t-$Cr_2O_3$.

### *3. Interfacial origin of magnetoresistance and anomalous Hall effect*

In this part we give some insight on the physical origin of the magnetotransport properties of Pt/t-$Cr_2O_3$ in the paramagnetic phase. As mentioned in the opening, anisotropic magnetoresistance related to magnetic proximity effect and spin-Hall magnetoresistance can be both observed at interfaces between Pt and ferromagnetic oxides [51].

It is well established that the contributions of SMR from AMR can be disentangled by exploiting their different symmetry with respect to current and field direction. [52] The results of the vectorial characterization of Pt/t-$Cr_2O_3$ at 200 K are reported in Figure 5(a).

Three non-overlapping MR curves are observed when the magnetic field **B** is applied parallel to the current (red dots), perpendicular in plane (black squares) or out of plane (blue triangles) indicating that both SMR and AMR are present. As a matter of fact, the longitudinal SMR is only sensitive to the in-plane component of magnetization perpendicular to the current ($\mathbf{B_{ip}} \perp \mathbf{j}$), [43] Since there is a difference between the cases of magnetic field applied parallel to the current ($\mathbf{B_{ip}} \parallel \mathbf{j}$) or out of the plane ($\mathbf{B_{oop}} \perp \mathbf{j}$) this represents the contribution of AMR.

The relative amplitude of the two effects demonstrates SMR to be the dominant term in the longitudinal resistivity being almost 5 times larger than AMR at 9 T. Coherently with what observed in Pt/YIG [51] spin-Hall related terms are prevalent over magnetic proximity effect in the change of longitudinal resistance.

Considering instead transverse resistivity $R_{xy}$, it is not possible to distinguish AHE coming from magnetic proximity or spin-Hall effect by symmetry reasons as also the spin-Hall magnetoresistance theory would predict a transverse resistance change linearly proportional to the out-of-plane component of magnetization. [43]

In reference [17], the anomalous part of the Hall effect in Pt/$Cr_2O_3$ was ascribed entirely to magnetic proximity effect. Indeed, the dominating role of MPE over SMR in the Hall signal was also demonstrated in a systematic study of interfaces between Pt and the ferrimagnetic insulator YIG [51].

Here we just give a qualitative discussion on the importance of MPE to explain the field amplitude dependence of the transverse resistivity in the paramagnetic phase of t-$Cr_2O_3$ (T = 200 K) Figure 5(b) shows the Hall signal as a function of the out-of-plane field in a range from -9 to +9 T. This large measurement range allows to better distinguish the linear ordinary Hall component to the non-linear part arising from interface effects. To allow for direct comparison, the linear slope due to ordinary effect is extrapolated as the red line. In this way it is easy to see that the superimposed anomalous term changes its sign at about |B|=200 mT. In SMR theory, the transverse signal should be directly proportional to the magnetization [43] and SMR alone cannot explain this change of sign. MPE instead, presents a complex and non-linear relationship between Hall signal and magnetization (possibly including sign change) as recently proposed in a theoretical model for AHE coming from interfaces. [47] Therefore, MPE can justify by itself this particular behavior. Finally, as SMR is indeed present in the system, it is also possible that both MPE and SMR could be observed in the transverse voltage with opposite-sign contributions. In this case, the change of sign of the anomalous Hall term with magnetic field would arise from the competition of the two effects depending on the relative strength and magnetic field dependence. Disentangling quantitatively SMR from MPE goes beyond the scope of this paper but remains clear that both phenomena play a role in the electronic transport at the Pt/t-$Cr_2O_3$ and neither of them should be a priori neglected in the study of similar interfaces.

## IV. CONCLUSIONS

In this work we discussed the magnetic properties of an uncommon phase of chromium (III) oxide, stabilized in the tetragonal phase by epitaxial growth on perovskite $BaTiO_3$, and the interfacial effects between this magnetic oxide and a neighboring Pt layer. The DFT shows that the t-$Cr_2O_3$ con be considered as a correlated compound with low connectivity and predicts an antiferromagnetic ground state. From electrical measurements, we show that sizeable magnetotransport effects both in longitudinal and transverse measurement configurations can be observed at this interface, both pointing out a magnetic phase transition at about 40 K. Signatures of magnetic proximity and spin-Hall magnetoresistance related to the high spin-orbit coupling in Pt are found. Despite the evidences of a proximity induced magnetic moment in Pt, we rule out that the resistance anomaly observed at low temperature originates from Kondo effect, coming instead from 2D weak localization in the ultrathin metal layer.

By validating the theoretical prediction of antiferromagnetism in t-$Cr_2O_3$, the approach here presented demonstrates the versatility of transport measurement as a tool for the investigation of new antiferromagnetic compounds. Finally, we have shown that, in general, electronic transport in metal/antiferromagnet interfaces cannot be simply described either by spin-Hall magnetoresistance theory or magnetic proximity effect alone. On the contrary, a comprehensive description can be achieved only if both phenomena are considered together.

## ACKNOWLEDGEMENTS

M.A. thanks prof. G. Jakob and prof. M. Kläui for allowing part of the cryogenic measurements and for fruitful discussion. This work was partially performed at Polifab, the micro and nanofabrication facility of Politecnico di Milano, and in the framework of the nanoscience foundry and fine analysis (NFFA-MIUR Italy) project. The work is supported by the Foundation for Polish Science through the International Research Agendas program cofinanced by the European Union within the Smart Growth Operational Programme. C.A. acknowledges the CINECA award under the ISCRA initiative IsC54 "CAMEO" and IsC69 "MAINTOP" Grant, for the availability of high performance computing resources and support. C.B., C.M. and S.P. acknowledge the support through grants FARB17PAGAN and FARB18CAVAL. S. Abate is acknowledged for his technical support.

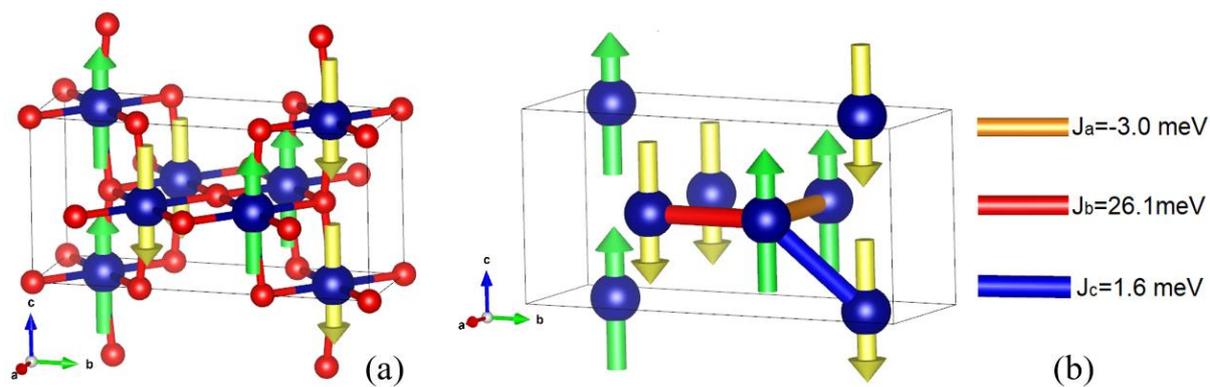

**Figure 1**: (a) Magnetic ground state from the DFT results. The up-spins (down-spins) are represented as green (yellow) arrows. The Cr and O atoms are represented as blue and red balls, respectively. (b) Magnetic exchanges $J_a$, $J_b$ and $J_c$ with their numerical values. For a better visualization, the oxygen atoms are not shown.

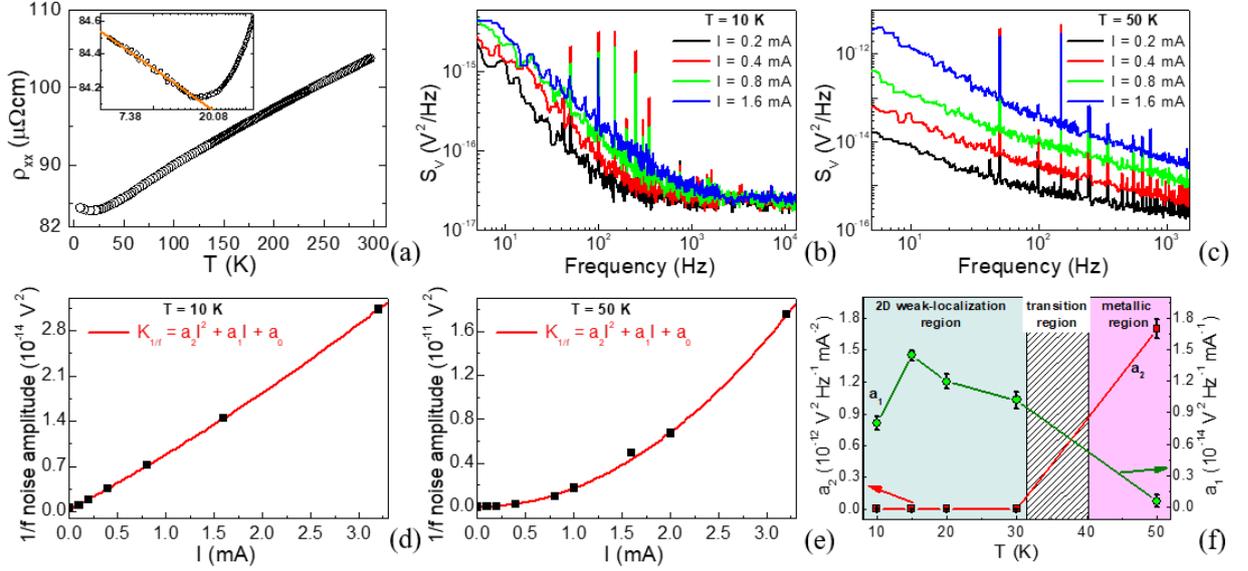

**Figure 2**: (a) Resistivity of Pt on CrOx as a function of temperature. Inset: zoom in the low-temperature region highlighting the logarithmic dependence of the resistance anomaly. The voltage-noise spectra are shown, for different bias currents, at temperatures of 10 K (b) and 50 K (c). For the same temperatures, it is also shown the current dependence of 1/f noise amplitude, at f = 90 Hz, (d) and (e), respectively. (f) The evidence of dominant quadratic (red squares) and linear (green circles) noise parameters is found in the metallic and in the 2D weak-localization region, respectively.

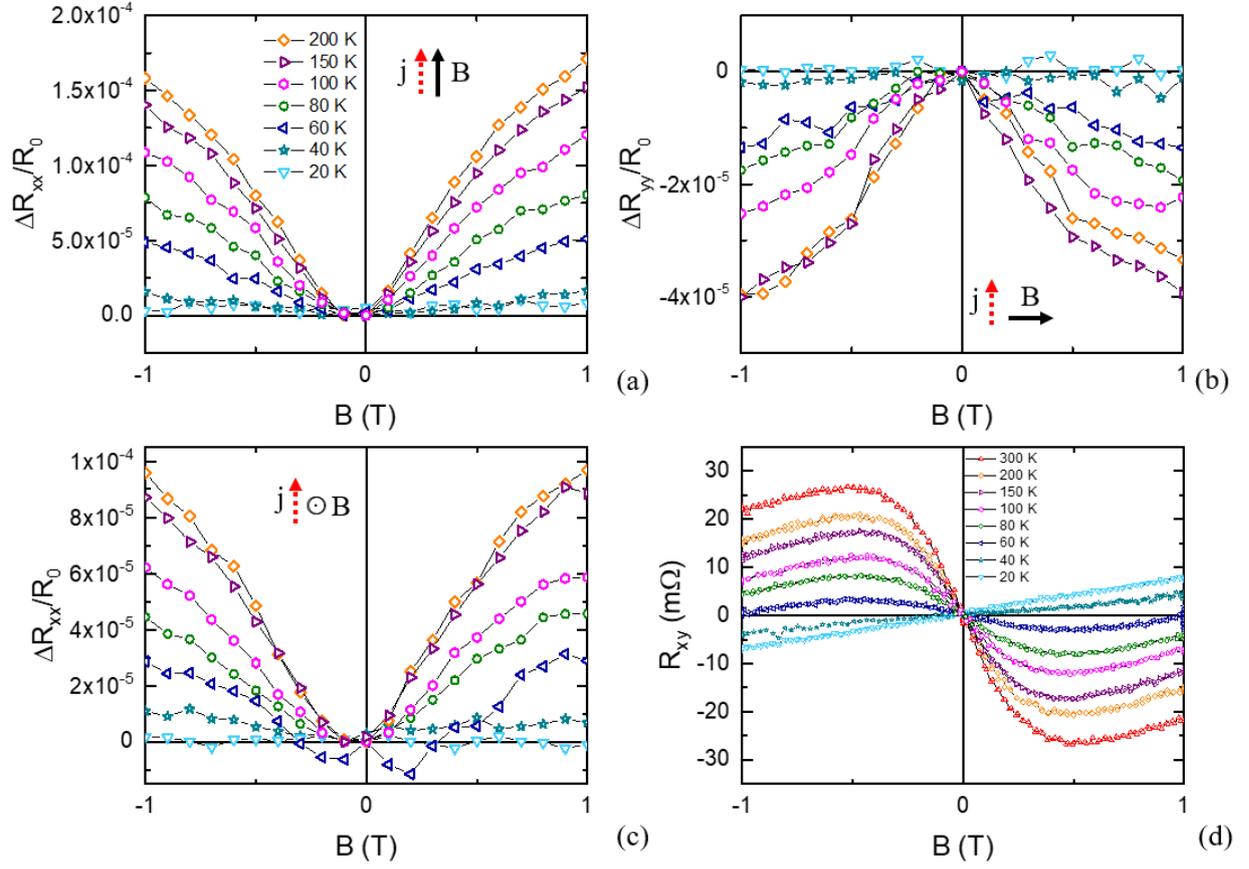

**Figure 3**: Magnetoresistance and Hall effect in Pt/t-Cr$_2$O$_3$ at different temperatures. (a) Magnetoresistance for field parallel to the current. (b) Field perpendicular to the current in plane. (c) Field out of plane. d) Hall effect.

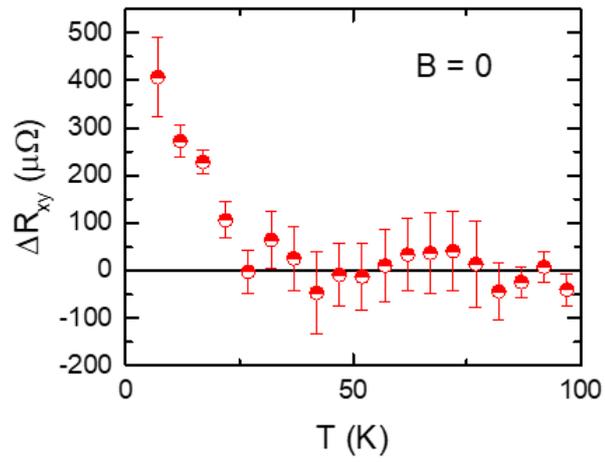

**Figure 4**: Difference in the residual anomalous Hall resistance at remanence between states set with an upward and a downward field cooling.

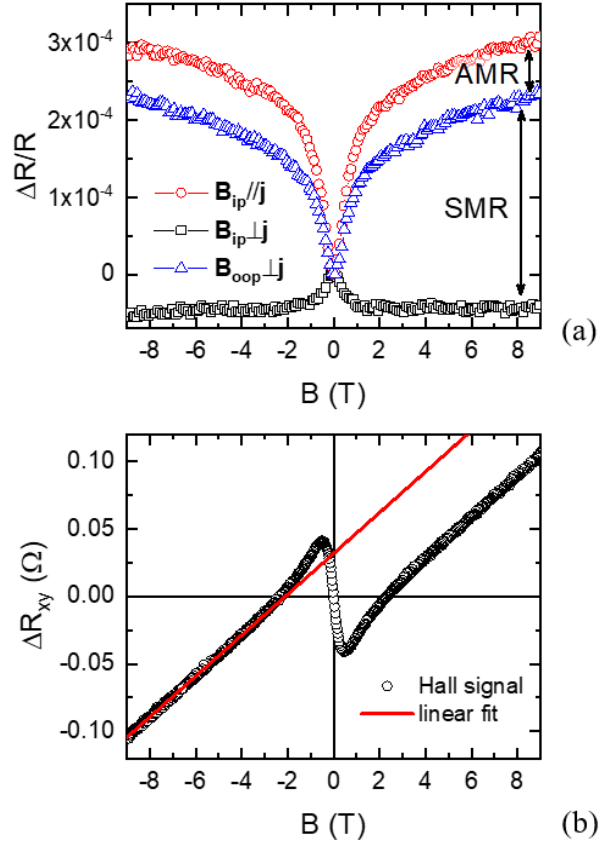

**Figure 5**: Contribution to magnetotransport of different interfacial effects at T=200 K. (a) vectorial characterization of longitudinal magnetoresistance highlighting both SMR and AMR. (b) Transverse resistance between -9 and +9 T. The red line is the linear fit of the experimental points between -9 T and -5 T representing the component due to ordinary Hall effect. Both a positive and a negative anomalous Hall component are observed.